\documentclass[reprint,3p,times]{elsarticle}
\usepackage{epstopdf}
\usepackage{graphicx}
\usepackage{algorithm}
\usepackage{algorithmic}
\usepackage{multirow}
\usepackage{amsmath}
\usepackage{amssymb}
\usepackage{amsthm}

\DeclareMathOperator*{\diag}{diag}

\DeclareMathOperator*{\minimize}{minimize}

\journal{Elsevier}

\begin{document}

\begin{frontmatter}
\title{Baselining Network-Wide Traffic by Time-Frequency Constrained Stable Principal Component Pursuit\tnoteref{label1}}
\tnotetext[label1]{This work is supported by the National Science Foundation of China (Nos. 61073013 and 90818024)
and the Open Project of State Key Laboratory of Software Development Environment (Nos. SKLSDE-2012ZX-15 and SKLSDE-2013ZX-34).}
\author[1]{Kai Hu}
\ead{hukai@buaa.edu.cn}
\author[1]{Zhe Wang\corref{cor1}}
\ead{wangzhe@cse.buaa.edu.cn}
\cortext[cor1]{Corresponding author.}
\author[1]{Baolin Yin}
\ead{yin@nlsde.buaa.edu.cn}
\address[1]{State Key Laboratory of Software Development Environment, Beihang University, Beijing 100191, China}

\begin{abstract}
The Internet traffic analysis is important to network management,
and extracting the baseline traffic patterns is especially helpful for some significant network applications.
In this paper, we study on the baseline problem of the traffic matrix satisfying a refined traffic matrix decomposition model,
since this model extends the assumption of the baseline traffic component to characterize its smoothness,
and is more realistic than the existing traffic matrix models.
We develop a novel baseline scheme,
named Stable Principal Component Pursuit with Time-Frequency Constraints (SPCP-TFC),
which extends the Stable Principal Component Pursuit (SPCP) by applying new time-frequency constraints.
Then we design an efficient numerical algorithm for SPCP-TFC.
At last, we evaluate this baseline scheme through simulations,
and show it has superior performance than the existing baseline schemes RBL and PCA.
\end{abstract}

\begin{keyword}
baseline of traffic matrix \sep
robust PCA \sep
time-frequency constraint \sep
numerical algorithm \sep
simulation.
\end{keyword}
\end{frontmatter}

\section{Introduction}
The Internet traffic analysis is of critical importance to network operation and management.
Usually, the total traffic is modeled by the superposition of diverse components corresponding to different user behaviors \cite{Lakhina_2004}, \cite{Lakhina_sigcomm}, \cite{Bandara_2011}, \cite{Wang_2012}.
The baseline traffic represents the most prominent traffic patterns \cite{Bandara_2011},
which is quite helpful for many significant network applications such as capacity planning, load balancing, and anomaly detection.
In the past, most studies devoted to estimating the trend of the single-link traffic \cite{Hajji_2003},
rather than extracting the common traffic pattern of the whole network,
however, the later is more informative for the manager of a large-scale network.
Recently, as the network-wide traffic measurement is becoming increasingly popular,
efficiently baselining the network-wide traffic turns into a practical and most urgent problem.

The traffic matrix a kind of network-wide traffic data,
and it represents the traffic exchanged between each Origin-Destination (OD) pair\footnote{The traffic traversing each OD pair is named an OD flow.} of the network.
Compared with other traffic data such as the link loads, it has a significant advantage \cite{Conti_2010}:
the OD flows are invariant under the changes of topology and routing.
Thus the traffic matrix shows the true intensity of the relationships between the OD pairs,
and hence is quite helpful for archiving the optimization in capacity planning and traffic engineering,
and detecting the network-wide anomalies more accurately.
The traffic matrix is obtained either by indirect estimation or by direct measurement \cite{Nie_2013}.
Until very recently, the estimation approach was still an active research topic.
But some things have changed around 2010 \cite{Willinger_2011},
because the netflow-enabled routers \cite{Cisco_2007} are increasingly deployed,
a large percentage of today's networks are able to measure themselves.
Hence in many cases, it is feasible to directly measure the traffic matrix now.

The baseline of a traffic matrix captures the common patterns among OD flows,
and it should be stable against the disturbance of anomaly traffic.
The Principal Component Analysis (PCA) was used for traffic matrix analysis first in \cite{Lakhina_2004},
and showed the low-rank nature of the baseline (i.e. the deterministic) traffic component,
but it performed poorly when the traffic matrix contains large anomalies \cite{Rubinstein_2009}, \cite{Wang_2012}.
Recently, the Robust Principal Component Analysis (RPCA) theory \cite{Ma_RPCA1},
which focus on recovering the low-rank matrix contaminated by the sparse matrix whose non-zeros entries may have large magnitudes,
has attracted wide attentions.
Candes et al. \cite{Ma_RPCA1} presented the Principal Component Pursuit (PCP) method,
and proofed that it could recover the low-rank matrix accurately under very board conditions.
Interestingly, the empirical characteristics of the traffic matrix are close to the structural hypotheses of the RPCA theory,
since the trends of different OD flows, which are highly correlated, fit for the low-rank hypothesis,
and the network anomalies, which rarely appear in time, fit for the sparse hypothesis.
Inspired by this fact, Abdelke et al. \cite{Abdelke_2010} first adopted the PCP method for network traffic analysis,
while their work mainly considered the anomaly detection problem.
Later, Bandara and Jayasumana \cite{Bandara_2011} proposed the Robust BaseLine (RBL) scheme,
which was also based on PCP and followed the exact "low-rank and sparsity" assumption,
and they argued that RBL performs better than several existing traffic baseline schemes.

Even so, it still makes sense to work on this topic more intensively.
In fact, the exact "low-rank and sparsity" traffic matrix model in \cite{Abdelke_2010} and \cite{Bandara_2011} is quite simple, and not very realistic.
On the one hand, considering the baseline time-series of each OD flow, as it represents the long-term and deterministic traffic trends such as the diurnal pattern, this time-series should be smooth enough. But this feature can not be characterized by the low-rank assumption.
On the other hand, the empirical OD flow traffic also contains the short-term fluctuations behavior with small magnitudes \cite{Soule_2007}.
In this case, the traffic matrix does not exactly meet the "low-rank and sparsity" assumption,
instead, it has a noise component.
In \cite{Wang_2012}, we modeled the noise traffic, but did not consider the smoothness of the baseline traffic.
Consequently, it is necessary to build a more realistic traffic matrix model, and consider the related traffic baseline problem.
In addition, the evaluations of traffic baseline schemes were not very sufficient in the previous studies. A key hurdle is that obtaining the ground-truth baseline of the real-world traffic matrix is impossible,
as a result, one could neither measure the accuracy of a baseline scheme,
nor compare different baseline schemes trustworthily.
Hence the simulation approach which contains the ground-truth information is needed in the evaluation process.

In this paper, we study on the baseline problem under a more realistic traffic matrix model,
and propose a novel baseline scheme to enforce the smoothness of the baseline traffic component.
Our contributions are listed as follows.
\begin{itemize}
  \item We present a refinement of the traffic matrix decomposition model in \cite{Wang_2012},
  which extends the descriptions of the baseline traffic to characterize its smoothness.
  \item We propose a novel traffic matrix baseline scheme named
  Stable Principal Component Pursuit with Time-Frequency Constraints (SPCP-TFC).
  As an extension of the Stable Principal Component Pursuit (SPCP) \cite{Zhou_2010},
  SPCP-TFC takes new time-frequency constraints.
  \item We design the Accelerated Proximal Gradient (APG) algorithm for SPCP-TFC, which has a fast convergence rate.
  \item We evaluate our baseline scheme through simulations and show it has superior performance than RBL and PCA.
\end{itemize}

\section{Methodology}
\subsection{A Refined Traffic Matrix Decomposition Model}
Suppose $X\in\mathbb{R}^{T\times P}$ is a traffic matrix,
and each column $X_{j}\in\mathbb{R}^{T}$ ($1\leq j\leq P$) is an OD flow in $T$ time intervals.
In \cite{Wang_2012}, we proposed the simple Traffic Matrix Decomposition Model (TMDM),
assuming $X$ is the sum of a low-rank matrix, a sparse matrix, and a noise matrix.
This model is equivalent to the data model of the generalized RPCA problem \cite{Zhou_2010},
and the low-rank deterministic traffic matrix corresponds to the baseline traffic\footnote{In the following discussion, the words "deterministic traffic" and "baseline traffic" are used interchangeably.}.
However, TMDM didn't consider the temporal characteristics of the baseline traffic.
Since the baseline traffic time-series of each OD flow represents the long-term and steady user behaviors,
it tends to display a smooth curve.
A number of mathematical tools, such as the wavelets and the splines \cite{Barford_2002}, can formulate smoothness.
As the most salient baseline traffic patterns are slow oscillation behaviors,
in this paper, we formulate the baseline traffic time-series as the sum of harmonics with low frequencies,
and thus establish a Refined Traffic Matrix Decomposition Model (R-TMDM):

\noindent\textbf{Definition 1} (R-TMDM)
\emph{The traffic matrix $X\in\mathbb{R}^{T\times P}$ is the superposition of
the deterministic (baseline) traffic matrix $A$, the anomaly traffic matrix $E$, and the noise traffic matrix $N$.
$A$ is a low-rank matrix,
and for each column time-series in $A$,
the Fourier spectra whose frequencies exceed a critical value $f_{c}$ are zeros;
$E$ is a sparse matrix with most entries being zeros,
but the non-zeros entries may have large magnitudes;
$N$ is a random noise matrix,
and each column time-series is a zero-mean stationary random process with a relatively small variance.}

As the OD flows in the backbone network are highly aggregated by superimposing independent traffic processes,
it is appropriate to model the noise traffic by the Gaussian processes following the central limitation theory \cite{Conti_2010}, \cite{Roughan_2002}.
For simplicity,
we assume each time-series $N_{j}$ ($1\leq j\leq P$) is the white Gaussian noise with variance $\sigma_{j}^{2}>0$ in this study\footnote{In future work, we plan to consider a more general model, i.e. each noise traffic time-series is a fractional Gaussian noise, whose temporal characteristics are more similar to the real-world backbone traffic.}.

\subsection{Stable Principal Component Pursuit with Time-Frequency Constraints}
Let $\|\cdot\|_{\ast}$, $\|\cdot\|_{1}$, and $\|\cdot\|_{F}$ denote the nuclear norm,
the $l_{1}$ norm, and the Frobenius norm, respectively.
The Stable Principal Component Pursuit (SPCP) method for the generalized RPCA problem solves this convex program \cite{Zhou_2010}:
\begin{equation}\label{SPCP}
\begin{split}
       & \ \minimize_{A, E, N}{\|A\|_{\ast}+\lambda\|E\|_{1}}\\
s. t. & \ \ A+E+N = X, \ \|N\|_{F}^{2}\leq\delta,
\end{split}
\end{equation}
where $\lambda > 0 $ is a balance parameter, and $\delta > 0$ is a constraint parameter.
The objective function of (\ref{SPCP}) combines the nuclear norm and the $l_{1}$ norm,
which are the convex relaxations of the rank function and the $l_{0}$ norm, respectively,
to enhance the low-rank structure of matrix $A$, as well as the sparsity of matrix $E$.

In this study, in order to extract the baseline traffic (i.e. matrix $A$) more accurately,
we extend SPCP by preserving its objective function and redesigning the constraint functions.
Firstly, considering the R-TMDM model,
it is necessary to add a constraint for the baseline traffic matrix based on its frequency-domain assumption.
Let $W = [W_{0} \cdots  W_{T-1}]_{T \times T}$ denote the discrete Fourier basis matrix of length $T$.
For each $0\leq k \leq T-1$, the Fourier basis $W_{k}$ is defined as
\begin{equation}
W_{k}(t) = \frac{1}{\sqrt{T}}e^{-i\frac{2\pi k}{T}(t-1)}, \ 1\leq t \leq T,
\end{equation}
with frequency $f_{k} = \min\{\frac{k}{T}, \frac{T-k}{T}\}$.
Suppose $W_{\mathrm{H}}$ is made up of the high-frequency bases $W_{k}$ in $W$ satisfying $f_{k} \geq f_{c}$,
and thus $W_{\mathrm{H}} (W_{\mathrm{H}})^{\top}$ is the projection operator to the high-frequency subspace.
Hence we add the following constraint for the baseline traffic:
\begin{equation}\label{baseline_constraint}
W_{\mathrm{H}} (W_{\mathrm{H}})^{\top}A =0^{T\times P}.
\end{equation}

Secondly, unlike the Frobenius norm inequality in (\ref{SPCP}),
we use a different constraint strategy for the noise traffic matrix $N$.
For each column vector $N_{j}$ ($1\leq j\leq P$),
consider its periodogram function $\{I_{N_{j}}(k)\}_{k=0}^{T-1}$:
\begin{equation}\label{periodogram}
I_{N_{j}}(k) = \left| W_{k}^{\top}N_{j} \right|^{2} =
\left|\frac{1}{\sqrt{T}}\sum_{t=1}^{T}N_{j}(t) e^{-i\frac{2\pi k}{T}(t-1)}\right|^{2}, \ 0\leq k \leq T-1.
\end{equation}
Let $\chi_{s}^2 \ (\chi_{s})$ denote the Chi-square (Chi) distribution with $s$ degrees of freedom.
It is well-known \cite{Kokoszka_2000} that
when $N_{j}$ are i.i.d. standard Gaussian variables\footnote{As the preprocessing for $X$,
for each column vector $X_{j}$ ($1\leq j\leq P$), we divide it by $\sigma_{j}$.
This operation normalizes the Gaussian variables in $N$,
and preserves all the hypotheses of $A$ and $E$ in the R-TMDM model. },
$\{2 I_{N_{j}}(k)\}_{k=1}^{T-1}$ are i.i.d $\chi_{2}^{2}$ variables,
and $I_{N_{j}}(0)$ is a $\chi_{1}^{2}$ variable.
Consequently,
$\{\sqrt{2}|W_{k}^{\top}N_{j}|\}_{k=1}^{T-1}$ are i.i.d $\chi_{2}$
variables,
and $|W_{0}^{\top}N_{j}|$ is a $\chi_{1}$
variable.
Using this property,
we design the time-frequency constraint for the noise traffic:
\begin{equation}\label{noise_constraint1}
\begin{cases}
\sqrt{2}\left\|[W_{1} \ \cdots \ W_{T-1}]^{\top}N_{j}\right\|_{\infty} \leq \delta_{1}, \ \left|W_{0}^{\top}N_{j}\right| \leq \delta_{2},  \ & (\mathrm{in \ the \ frequency \ domain})\\
\left\|N_{j}\right\|_{\infty} \leq \delta_{3},     \ & (\mathrm{in \ the \ time \ domain})\\
\end{cases}
\ 1\leq j \leq P,
\end{equation}
where $\delta_{1},\delta_{2},\delta_{3} > 0$ are positive parameters.
Equivalently, $N$ lies in the convex subset $B(\delta_{1},\delta_{2},\delta_{3})$:
\begin{equation}\label{noise_constraint2}
B(\delta_{1},\delta_{2},\delta_{3}) \triangleq \left\{N \
\begin{array}{|l}
N\in\mathbb{R}^{T\times P} \ \mathrm{and} \ \mathrm{satisfies} \ (\ref{noise_constraint1}) \ \mathrm{under} \ \mathrm{the} \ \mathrm{parameters} \ \delta_{1}, \delta_{2}, \delta_{3} > 0
\end{array}
\right\}.
\end{equation}
Here we explain the main idea of (\ref{noise_constraint1}).
Following the R-TMDM model,
the frequency-domain distribution of the baseline traffic is highly concentrated.
Consequently, if the noise traffic is mixed with plenty of baseline traffic,
some of its low-frequency spectra tends to be enlarged dramatically and can be detected easily.
Thus, using the frequency-domain constraint in (\ref{noise_constraint1}),
which utilizes the distribution properties of the periodogram function in (\ref{periodogram}),
we could effectively eliminate the baseline traffic from the noise traffic.
Meanwhile, the time-domain constraint in (\ref{noise_constraint1}) controls the magnitudes of the noise traffic and eliminates the large anomaly traffic.
However, the Frobenius norm constraint in (\ref{SPCP}) is harder to filter the baseline traffic from the noise traffic,
since in the time domain, the baseline traffic is not highly concentrated.

Lastly, using the new constraints (\ref{baseline_constraint}) and (\ref{noise_constraint1}),
we propose our traffic baseline scheme,
named Stable Principal Component Pursuit with Time-Frequency Constraints (SPCP-TFC),
which outputs $A$ as the baseline traffic matrix:
\begin{equation}\label{SPCP-TFC}
\begin{split}
      & \ \minimize_{A, E, N}{\|A\|_{\ast}+\lambda\|E\|_{1}} \\
s.t. & \  \ A+E+N=X, \  W_{\mathrm{H}} (W_{\mathrm{H}})^{\top}A =0^{T\times P},  \
\mathcal{I}_{B(\delta_{1},\delta_{2},\delta_{3})}(N) = 0,
\end{split}
\end{equation}
where $\mathcal{I}_{B(\delta_{1},\delta_{2},\delta_{3})}(N)$ denotes the indicator function\footnote{For each set $C\subseteq\mathbb{R}^{T\times P}$,
the indicator function $\mathcal{I}_{C}(\cdot): \mathbb{R}^{T\times P} \rightarrow \{0, +\infty\}$ is defined by:
\begin{equation*}
\mathcal{I}_{C}(N) =
\begin{cases}
0           &\mathrm{if} \ N\in C,\\
+\infty &\mathrm{otherwise}.
\end{cases}
\end{equation*}
} of $B(\delta_{1},\delta_{2},\delta_{3})$.
In this study, we set $\delta_{1} = 3.03$, $\delta_{2}=\delta_{3} = 2.56$,
corresponding to the $99\%$ confidence intervals of the $\chi_{2}$, $\chi_{1}$, and Gaussian variables, respectively,
and it makes $B(\delta_{1},\delta_{2},\delta_{3})$ contain the main power of $N$.

\subsection{Numerical Algorithm}

SPCP-TFC solves a constrained program which is computational expensive.
Applying the numerical techniques introduced in \cite{Lin_2009} to accelerate the convergence,
we derive the Accelerated Proximal Gradient (APG) algorithm for SPCP-TFC,
which considers this relaxed unconstrained program:
\begin{equation}\label{relaxation}
\begin{split}
& \minimize_{A, E, N} F(A,E,N) \triangleq \mu g(A,E,N) + f(A,E,N) \\
& f(A,E,N) \triangleq \frac{1}{2}\|X-A-E-N\|_{F}^{2} + \frac{\beta}{2}\|W_{\mathrm{H}} (W_{\mathrm{H}})^{\top}A\|_{F}^{2}, \\
& g(A,E,N) \triangleq \|A\|_{\ast} + \lambda\|E\|_{1} + \mathcal{I}_{B(\delta_{1},\delta_{2},\delta_{3})}(N). \\
\end{split}
\end{equation}
The function $f(A,E,N)$, which is convex and differentiable, penalizes violations of the two equality constraints in (\ref{SPCP-TFC}),
and $\beta>0$ is a balancing parameter.
$g(X,E,N)$ is a linear combination of three convex but non-differentiable functions.
$\mu > 0$ is a relaxation parameter, as $\mu\searrow 0$,
the solution to (\ref{relaxation}) can closely approximate the solution to (\ref{SPCP-TFC}).
$\nabla f$ denotes the gradient of $f$, which equals to
\begin{equation*}\label{gradient}
\nabla f(A,E,N) =
\left[
  \begin{array}{c}
    [I+\beta W_{\mathrm{H}} (W_{\mathrm{H}})^{\top}]A + E+N-X \\
    A+E+N-X \\
    A+E+N-X \\
  \end{array}
\right]
\end{equation*}
and is Lipschitz continuous by the following proposition:

\noindent\textbf{Proposition 1}
\emph{$\nabla f(A,E,N)$ is Lipschitz continuous,
and the Lipschitz constant is $L = \sqrt{9 + 4\beta + \beta^{2}}$.}

The proof is shown in the appendix section.

\quad


Instead of directly minimizing $F(A,E,N)$, the APG algorithm minimizes a sequence of quadratic approximations,
denoted as $Q(A,E,N,Y^{A},Y^{E},Y^{N})$, of $F(A,E,N)$ at a specially chosen point $(Y^{A},Y^{E},Y^{N})$
(renewed in each step):
\begin{equation}
\begin{split}
Q(A,E,N,Y^{A},Y^{E},Y^{N}) \triangleq & \mu g(A,E,N)  + f(Y^{A},Y^{E},Y^{N}) +
\left\langle \nabla  f(Y^{A},Y^{E},Y^{N}), (A,E,N)-(Y^{A},Y^{E},Y^{N}) \right\rangle + \\
                                                                     & \frac{L}{2}\left\| (A,E,N) - (Y^{A},Y^{E},Y^{N}) \right\|_{F}^{2}. \\
\end{split}
\end{equation}
It can be derived that
\begin{equation}\label{APG_subproblem}
\begin{split}
   & \minimize_{A,E,N} Q(A,E,N,Y^{A},Y^{E},Y^{N})\\
= & \minimize_{A} \left\{ \mu\|A\|_{\ast}  + \frac{L}{2}\|A-G^{A} \|_{F}^{2} \right\} + \\
   & \minimize_{E} \left\{ \mu\lambda\|E\|_{1}  + \frac{L}{2}\|E-G^{E} \|_{F}^{2} \right\} + \\
   & \minimize_{N} \left\{\mu\mathcal{I}_{B(\delta_{1},\delta_{2},\delta_{3})}(N)  + \frac{L}{2}\|N-G^{N} \|_{F}^{2} \right\}  + \mathrm{Constant} \\
\end{split}
\end{equation}
where
\begin{equation*}
\begin{cases}
G^{A} = Y^{A} - \frac{1}{L}\left( [I + \beta W_{\mathrm{H}} (W_{\mathrm{H}})^{\top}]Y^{A} + Y^{E} + Y^{N} - X \right)\\
G^{E} = Y^{E} - \frac{1}{L}\left(Y^{A} + Y^{E} + Y^{N} - X\right)\\
G^{N} = Y^{N} - \frac{1}{L}\left(Y^{A} + Y^{E} + Y^{N} - X\right)
\end{cases}
\end{equation*}
Thus problem (\ref{APG_subproblem}) splits into three subproblems,
which are equivalent to the \emph{proximity operators} associated with the convex functions
$\frac{\mu}{L}\|A\|_{\ast}$,
$\frac{\mu\lambda}{L}\|E\|_{1}$,
and $\frac{\mu}{L}\mathcal{I}_{B(\delta_{1},\delta_{2},\delta_{3})}(N)$, respectively.
Using the basic results in \cite{Combettes_2010}, they have explicit solutions as follow:
\begin{equation}
\arg\min_{A} \left\{ \mu\|A\|_{\ast}  + \frac{L}{2}\|A-G^{A}\|_{F}^{2} \right\} =
U\mathcal{S}_{ \frac{\mu}{L} }[\Sigma]V^{\top},
\end{equation}
\begin{equation}
\arg\min_{E} \left\{ \mu\lambda\|E\|_{1}  + \frac{L}{2}\|E-G^{E} \|_{F}^{2} \right\} =
\mathcal{S}_{\frac{\mu\lambda}{L}}[G^{E}],
\end{equation}
\begin{equation}
\arg\min_{N}\left\{ \mu\mathcal{I}_{B(\delta_{1},\delta_{2},\delta_{3})}(N)  + \frac{L}{2}\|N-G^{N} \|_{F}^{2}\right\} = \mathcal{P}_{B(\delta_{1},\delta_{2},\delta_{3})}[G^{N}],
\end{equation}
where $U\Sigma V^{\top}$ is the singular value decomposition (SVD) of $G^{A}$,
$\mathcal{P}_{B(\delta_{1},\delta_{2},\delta_{3})}$ is the projection operator to $B(\delta_{1},\delta_{2},\delta_{3})$,
and $\mathcal{S}_{\epsilon}$ is the soft-thresholding operator with parameter $\epsilon>0$ \cite{Wang_2012}.
For each $X\in\mathbb{R}^{T\times P}$,
 $\mathcal{S}_{\epsilon}[X] \in\mathbb{R}^{T\times P} $ and it satisifies
$$\mathcal{S}_{\epsilon}[X](i,j) \triangleq \mathrm{sign}(X(i,j))\max\{|X(i,j)|-\epsilon,0\}, \ 1\leq i\leq T, \ 1\leq j\leq P.$$

\begin{algorithm}[tbhp]
\caption{APG for SPCP-TFC}\label{algo1}
\textbf{Input}: traffic matrix $X\in \mathbb{R}^{T\times P}$.

\textbf{Normalization:} $X=X/ \diag\{\sigma_{j}\}$.

\textbf{Initialization:} $A_{0}=A_{-1}=E_{0}=E_{-1}=N_{0}=N_{-1}=0$; $t_{0}=t_{-1}=1$; $k=0$.
$\mu_{0}=0.99\|X\|_{2}$; $\overline{\mu}=10^{-5}\mu_{0}$; $\eta=0.9$.

$L = \sqrt{9+4\beta+\beta^{2}}$; $\lambda = 1/\sqrt{\max(T, P)}$.

\textbf{While} \ not converged \textbf{do}

\quad $Y_{k}^{A}=A_{k}+\frac{t_{k-1}-1}{t_{k}}(A_{k}-A_{k-1})$;

\quad $Y_{k}^{E}=E_{k}+\frac{t_{k-1}-1}{t_{k}}(E_{k}-E_{k-1})$;

\quad $Y_{k}^{N}=N_{k}+\frac{t_{k-1}-1}{t_{k}}(N_{k}-N_{k-1})$;

\quad $G_{k}^{A}=Y_{k}^{A}-\frac{1}{L}\left([I +\beta W_{\mathrm{H}} (W_{\mathrm{H}})^{\top}]Y_{k}^{A}+Y_{k}^{E}+Y_{k}^{N}-X\right)$;

\quad $G_{k}^{E}=Y_{k}^{E}-\frac{1}{L}\left(Y_{k}^{A}+Y_{k}^{E}+Y_{k}^{N}-X\right)$;

\quad $G_{k}^{N}=Y_{k}^{N}-\frac{1}{L}\left(Y_{k}^{A}+Y_{k}^{E}+Y_{k}^{N}-X\right)$;

\quad $(U,S,V)=\mathrm{svd}[G_{k}^{A}]$; \ \ // Singular Value Decomposition.

\quad $A_{k+1}=U\mathcal{S}_{\frac{\mu}{L}}[S]V^{\top}$;

\quad $E_{k+1}=\mathcal{S}_{\frac{\lambda\mu}{L}}[G_{k}^{E}]$;

\quad $N_{k+1}=\mathcal{P}_{B(\delta_{1},\delta_{2},\delta_{3})}[G_{k}^{N}]$;

\quad $t_{k+1}=(1+\sqrt{4t_{k}^{2}+1})/2$;

\quad $\mu_{k+1}=\max(\eta\mu_{k},\overline{\mu})$;

\quad $k=k+1$.

\textbf{End while}

\textbf{Output}: $A=A_{k}\cdot \diag\{\sigma_{j}\}$ is the baseline traffic matrix.
\end{algorithm}
Based on these discussions,
we present the APG algorithm for SPCP-TFC in Algorithm \ref{algo1},
and some of its initializations are borrowed from \cite{Lin_2009}.
The setting of $\beta$ is discussed in Section 3, actually, our experiment result is stable in a large range of $\beta$ values.
Moreover, the following theorem proves that Algorithm \ref{algo1} has the $O(1/k^{2})$ convergence rate.
Since the proof of this theorem is quite close to Theorem 4.4 in \cite{Beck_2009},
we omit it and directly summarize this result:

\noindent\textbf{Theorem 1}
\emph{Let $F(A,E,N)=\overline{\mu}g(A,E,N)+f(A,E,N)$.
Then for all $k>k_{0}=\left\lceil\log\left(\frac{\mu_{0}}{\overline{\mu}}\right)/\log\left(\frac{1}{\eta}\right)\right\rceil$,
we have
\begin{equation}
F(X_{k})-F(X^{\ast})\leq \frac{2\sqrt{9 + 4\beta + \beta^{2}}\left\|X_{k_{0}}-X^{\ast}\right\|_{F}^{2}}{(k-k_{0}+1)^{2}},
\end{equation}
where $X_{k}=(A_{k},E_{k},N_{k})$ is defined in Algorithm \ref{algo1},
and $X^{\ast}=(A^{\ast},E^{\ast},N^{\ast})$ is a solution to (\ref{relaxation}) when $\mu=\overline{\mu}$.}

\section{Simulation Results}\label{simulation_results}
We evaluate the SPCP-TFC scheme by synthetic traffic matrices,
because the simulation approach could
provide the ground-truth baseline data,
perform multiple experiments independently,
and vary simulation parameters for the sensitivity analysis.
Consequently, we first introduce the simulation techniques in this section.

Suppose the backbone network we studied has $10$ nodes,
each independent traffic measurement process lasts for one week,
and the minimal time interval is $\Delta t$ = $5$ minutes.
Under these settings,
each synthetic traffic matrix $X\in\mathbb{R}^{T\times P}$ records $P$ = $10^{2}$ = $100$ OD flows' traffic
during $T$ = $7\times 24\times 12$ = $2016$ consecutive time intervals.
In our simulation, $X$ is is made up of the following three components,
and the baseline component should be generated first since the others need some of its settings.
\begin{itemize}
  \item The baseline traffic matrix $A$.
  Each column $A_{j}$ $(1\leq j\leq P)$ is the sum of a positive constant $a_{j,0}$ and some sine functions:
  \begin{equation}\label{sine_function}
  A_{j}(t) = a_{j,0} + \sum_{m=1}^{M}a_{j,m}\sin\left(\frac{2\pi l_{m} t}{T} +\varphi_{j,m}\right),
  \ 1\leq t\leq T.
  \end{equation}
  In this study, we mainly use the sine function to simulate the periodic patterns because it is the most commonly considered function to capture the trend of the Internet traffic \cite{Lakhina_2004}, \cite{Lakhina_sigcomm}, \cite{Rubinstein_2009}, meanwhile, we will consider other periodic function at the end of this section.
  The mean values $\{a_{j,0}\}_{j=1}^{P}$ of different columns satisfies the well-known gravity model \cite{Roughan_2005},
  and in each simulation run, their sum is fixed at the constant $10^{6}$.
  As the period of the $m$-th sine is $\left(\frac{T}{l_{m}}\times\frac{\Delta t}{60}\right)$ hours,
  we choose $\{l_{m}\}$ = $\{7, 14, 28, 56, 112\}$,
  to simulate the $\{24, 12, 6, 3, 1.5\}$ hours periodical traffic patterns\footnote{This choice is based on the empirical studies \cite{Lakhina_2004}\cite{Rubinstein_2009}, for simplicity, here we neglect the periodical traffic patterns longer than 24 hours.}.
  In addition, the amplitudes of these sine functions decay quickly as $m$ increases by $a_{j,m+1}$ = $0.5 a_{j,m}$,
  and the phase parameters $\{\varphi_{j,m}\}$ are independently and uniformly sampled from the interval $[-\frac{\pi}{5}, \frac{\pi}{5}]$.
  \item The anomaly traffic matrix $E$.
  In each simulation run,
  the non-zero entries (each corresponds to an anomaly incident) of $E$ takes $1\%$ randomly chosen positions,
  and in each column $E_{j}$ $(1\leq j\leq P)$, the volume of an anomaly is fixed at $0.8a_{j,0}$.
  \item The noise traffic matrix $N$.
  Each column $N_{j}$ $(1\leq j\leq P)$ is generated by $T$ independent Gaussian variables $N(0,\sigma_{j}^{2})$,
  where $\sigma_{j}=\alpha a_{j, 0}$, and the parameter $\alpha>0$ controls the noise rate.
\end{itemize}
We first conduct $100$ baseline traffic matrices simultaneously,
then for each baseline matrix, generate one anomaly traffic matrix and two noise traffic matrices ,
($\alpha$ is chosen as $0.1$ and $0.2$, simulating a low-level noise condition and a high-level noise condition, respectively).
Consequently, $200$ synthetic traffic matrices are provided to our experiment.
We verified that all these matrices satisfy the hypotheses of the R-TMDM model,
the rank of each baseline matrix is $11$,
and its critical frequency $f_{c}$ = $\frac{112}{2016}$.

In this study, SPCP-TFC is compared with other two network-wide traffic baseline schemes:
RBL \cite{Bandara_2011} and PCA \cite{Lakhina_2004}.
For the PCA scheme, the number of principal components is fixed at $11$,
which is equal to the rank of the ground-truth baseline matrix.
We applying these baseline schemes independently to capture the baseline of the synthetic traffic matrices,
and evaluate their results using three distinct metrics.

\begin{figure}[!htb]
\centering
\scalebox{0.65}[0.65]{\includegraphics*{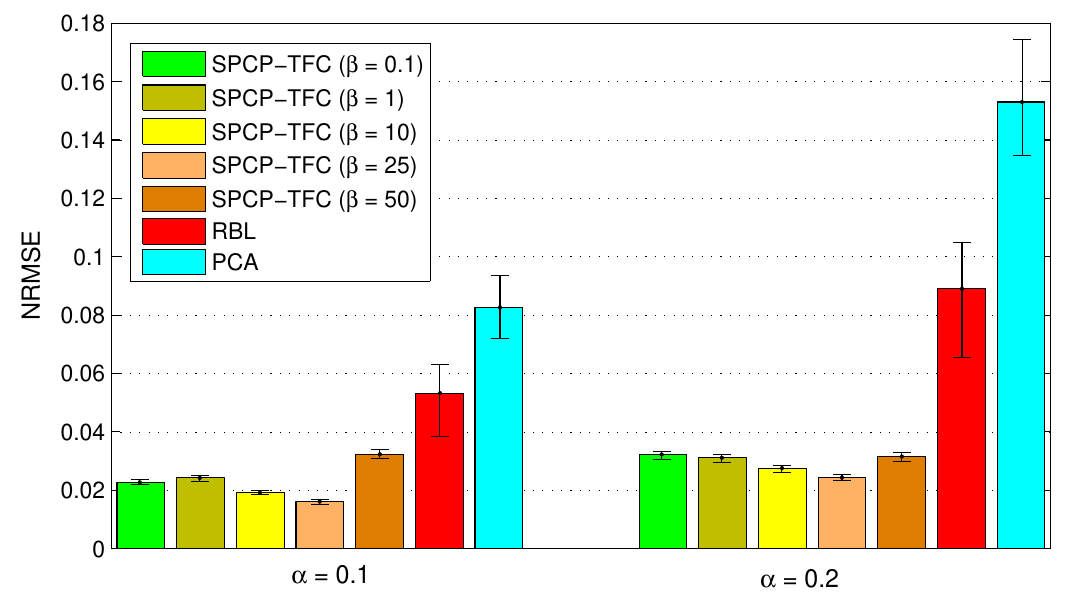}}
\begin{center}
\caption{The NRMSEs between the resulting and the ground-truth baseline traffic matrices.
The bar shows the median and the error bar shows 10\% and 90\% percentile.
Left group: $\alpha$ = $0.1$; right group: $\alpha$ = $0.2$. }\label{NRMSE}
\end{center}
\vspace{10pt}
\end{figure}
The first metric is the \emph{Normalized Root Mean Squared Error} (NRMSE)
between the ground-truth baseline traffic matrix $A$ and its estimation $\widehat{A}$:
\begin{equation}
\mathrm{NRMSE} = \frac{\|A-\widehat{A}\|_{F}}{\|A\|_{F}} .
\end{equation}
Fig. \ref{NRMSE} illustrates the NRMSEs of different baseline schemes under two noise levels.
For SPCP-TFC, we test different values of parameter $\beta\in\{0.1, 1, 10, 25, 50\}$.
It can be observed that,
SPCP-TFC archives significant lower NRMSEs than RBL and PCA,
and this result is stable in a large range of $\beta$ values.
Moreover, our traffic baseline scheme shows the best performance when $\beta$ = $25$:
under the low (high) noise level,
the median of NRMSEs for SPCP-TFC is as $30.3\% (27.3\%)$ as that for RBL,
and is as $19.5\% (15.9\%)$ as that for PCA.
Hence, we only consider SPCP-TFC with this fixed value in the following discussions.

\begin{figure}[!htb]
\centering
\scalebox{0.6}[0.6]{\includegraphics*{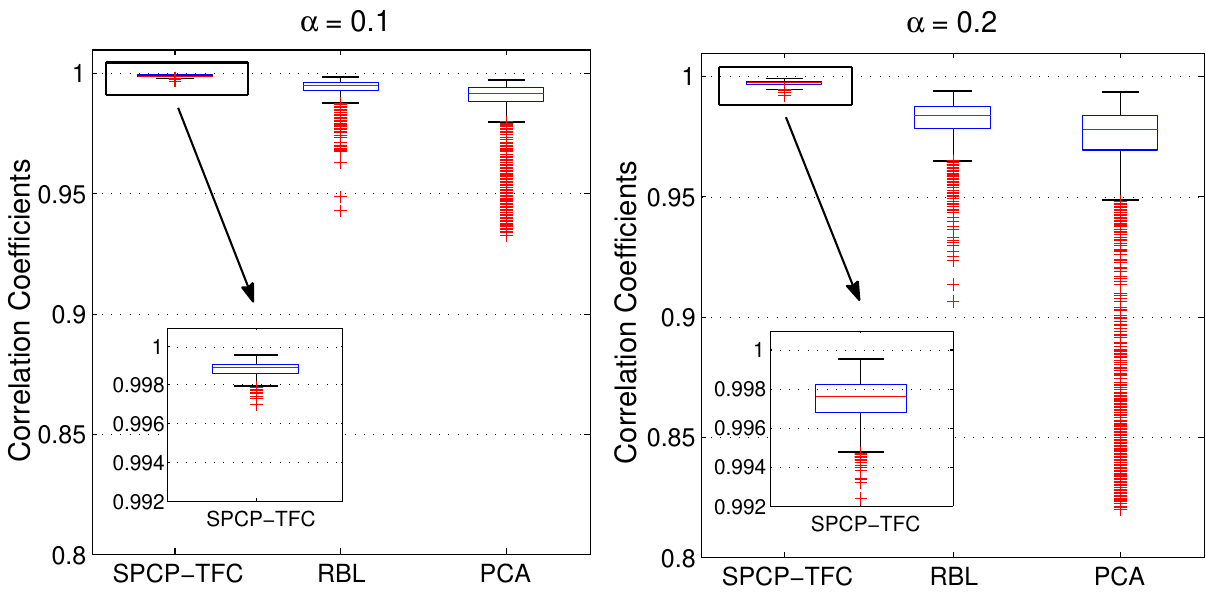}}
\begin{center}
\caption{Boxplots of the Pearson Correlation Coefficients between resulting and ground-truth baseline traffic flows.
The results of SPCP-TFC are enlarged in small plots.
Left panel: $\alpha$ = $0.1$; right panel: $\alpha$ = $0.2$.}\label{Correlation}
\end{center}
\vspace{10pt}
\end{figure}
The second metric characterizes the temporal correlation between each pair of ground-truth and resulting baseline traffic flows
using their \emph{Pearson Correlation Coefficient}.
As a traffic matrix contains 100 flows, under each noise level,
we compute $100 \times 100$ = $10^{4}$ correlation coefficients for each baseline scheme.
We display boxplots of these coefficients in Fig. \ref{Correlation}.
It is shown that the correlation coefficients for SPCP-TFC are very close to one,
and have an obviously greater median than RBL and PCA.
Thus SPCP-TFC extracts the temporal characteristics of the baseline traffic more precisely.

\begin{figure}[!htb]
\centering
\scalebox{0.6}[0.6]{\includegraphics*{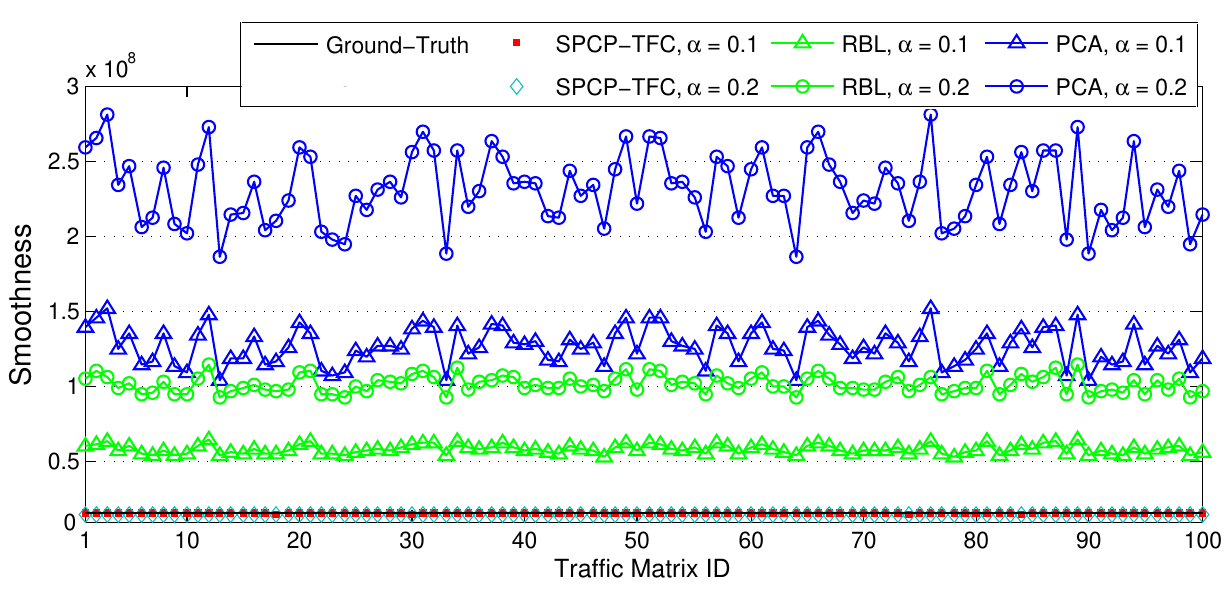}}\\
\vspace{5pt}
\scalebox{0.6}[0.6]{\includegraphics*{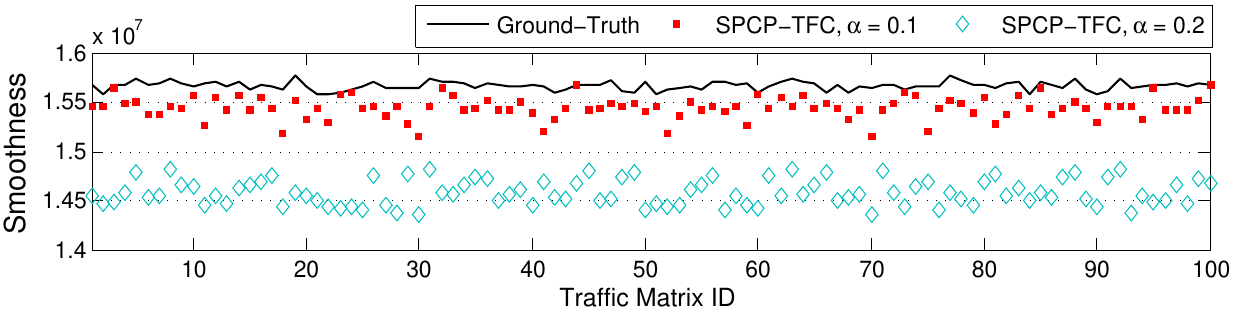}}
\begin{center}
\caption{The smoothness of resulting and ground-truth baseline traffic matrices.
An enlarged view of SPCP-TFC's smoothness is shown in the bottom panel.}\label{Smoothness}
\end{center}
\vspace{10pt}
\end{figure}

\begin{table}[!htb]
\begin{tabular*}{\columnwidth}{@{}lllll}
\multicolumn{5}{l}{\footnotesize \textbf{Table 1}}\\
\multicolumn{5}{l}{\footnotesize A comparison between three different baseline schemes on the normalized mean smoothness values}\\
\hline
\footnotesize $\alpha$  &  \footnotesize SPCP-TFC & \footnotesize RBL & \footnotesize PCA & \footnotesize Ground-Truth \\
\hline
\footnotesize 0.1  & \footnotesize \textbf{0.99} & \footnotesize 3.72 & \footnotesize 8.10   & \footnotesize 1.00 \\
\footnotesize 0.2  & \footnotesize \textbf{0.95} & \footnotesize 6.48 & \footnotesize 14.78 & \footnotesize 1.00 \\
\hline
\end{tabular*}
\vspace{10pt}
\end{table}
The third metric characterizes baseline traffic on smoothness,
which adds up the \emph{Total Variations} of the traffic flows in a baseline traffic matrix $A$ (ground-truth or estimated):
\begin{equation}
\mathrm{Smoothness}(A) \triangleq \sum_{j=1}^{P}\sum_{t=1}^{T-1}\left|A_{j}(t+1)-A_{j}(t)\right|.
\end{equation}
A small value of this metric would indicate that the baseline traffic is sufficiently smooth.
Applying this metric to the aforementioned baseline results,
we display the curve of 100 smoothness values (arranged by traffic matrix ID) for each baseline scheme,
and for the 100 ground-truth values, in Fig. \ref{Smoothness}.
We then compute the mean smoothness of each baseline scheme,
and normalize it by dividing the mean smoothness of the ground-truth baseline traffic matrices.
These normalized mean smoothness values (under two noise levels) are summarized in Tab. 1.
The estimated baseline traffic matrices by RBL and PCA are significantly more coarse than the ground-truth baseline traffic,
and their mean smoothness values are larger than three times and eight times of the ground-truth value, respectively.
Instead, SPCP-TFC leads to more accurate baseline estimations on smoothness.
This is because the high-frequency traffic,
which has larger total variation,
can be successfully eliminated from the baseline by SPCP-TFC.

From Fig. \ref{NRMSE}-\ref{Smoothness} and Tab. 1,
we can also observe that when the noise rate $\alpha$ grows from $0.1$ to $0.2$,
the performance of our baseline scheme shows a decline:
the NRMSEs raise, the correlation coefficients drop down,
and the smoothness values of resulting baseline traffic matrices diverge from the ground-truth more evidently.
This is mainly because the low-frequency component of the noise traffic,
which could not be directly distinguished from the baseline traffic, is proportional to $\alpha$.

\begin{figure}[!htb]
\centering
\scalebox{0.65}[0.65]{\includegraphics*{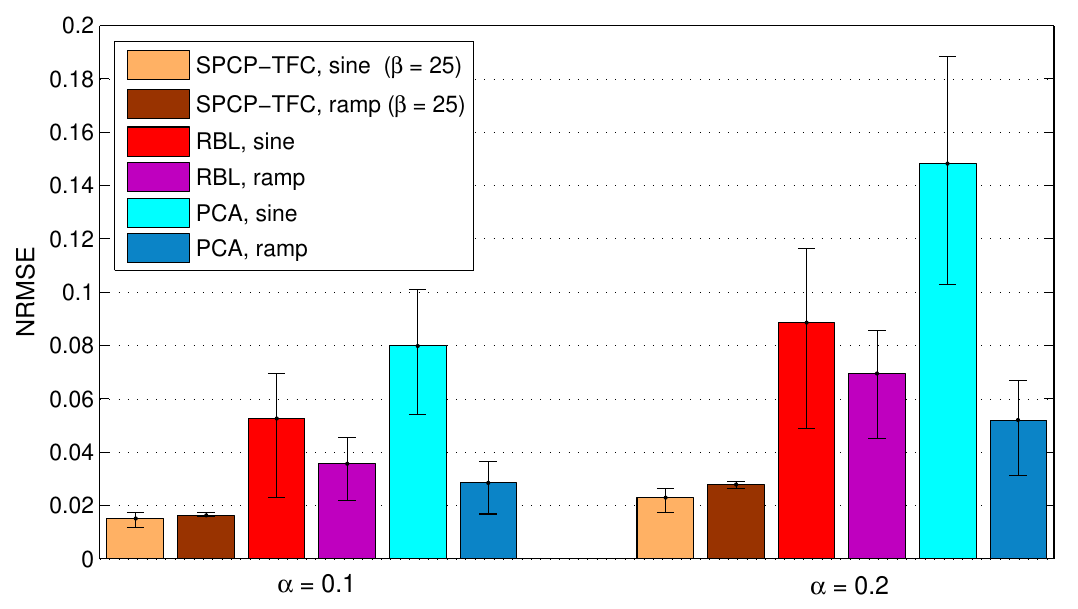}}
\begin{center}
\caption{The NRMSEs between the resulting and the ground-truth baseline traffic matrices,
both the "sine" dataset and the "ramp" dataset are experimented.
The bar shows the median and the error bar shows 10\% and 90\% percentile.
Left group: $\alpha$ = $0.1$; right group: $\alpha$ = $0.2$. }\label{NRMSE_ramp}
\end{center}
\vspace{10pt}
\end{figure}
As the ending part of this section, we evaluate traffic baseline methods by an additional simulation experiment in which the periodic traffic patterns do not follow the sine functions. In these traffic matrices, the sine functions are replaces with periodic ramp functions\footnote{We refer these new-generated traffic matrices as the "ramp" dataset, and the aforementioned ones as the "sine" dataset},
for instance, the diurnal traffic
\footnote{Similarly, other periodic traffic patterns are generated by stretching $\mathrm{ramp}(t)$.}
originally represented by $\sin(\frac{2\pi l_{1}t}{T}+ \varphi_{j,1})$ in (\ref{sine_function}) is simulated by
\begin{equation}\label{ramp_function}
\mathrm{ramp}(t)\triangleq
\begin{cases}
8t/T       &\mathrm{if} \ \ 0    \leq (t \ \% \ T) < T/8, \\
1          &\mathrm{if} \ \ T/8  \leq (t \ \% \ T) < 3T/8, \\
4 - 8t/T   &\mathrm{if} \ \ 3T/8 \leq (t \ \% \ T) < 5T/8, \\
-1         &\mathrm{if} \ \ 5T/8 \leq (t \ \% \ T) < 7T/8, \\
-8 + 8t/T  &\mathrm{otherwise}.
\end{cases}
\end{equation}
where $(a \ \% \ b)$ means the remainder of dividing (integer) $a$ by (integer) $b$.
We demonstrate SPCP-TFC, RBL, and PCA on this "ramp" dataset which has the same number of samples as the "sine" dataset,
and exhibit the NRMSEs of the three baseline schemes under two noise levels ($\alpha = 0.1, 0.2$) in Fig. \ref{NRMSE_ramp}.
As a comparison, the NRMSEs of these baseline schemes in the "sine" dataset experiment (Fig. \ref{NRMSE}) are displayed together.
For SPCP-TFC, we just depict the results when the balancing parameter $\beta = 25$,
because this value leads to the minimal errors both in the "sine" experiment and the "ramp" experiment.
We discover the following two features in Fig. \ref{NRMSE_ramp}:
\begin{itemize}
\item For SPCP-TFC, the median of NRMSEs shows a small increase ($7.9\%$ and $21.8\%$ when $\alpha$ equals $0.1$ and $0.2$, respectively) when we change the "sine" dataset as the "ramp" dataset in experiment;
\item Compared with RBL and PCA, our baseline method still has relatively lower NRMSEs on the "ramp" dataset.
Under the low(high) noise level, the median of NRMSEs for SPCP-TFC is as $42.1\%(40.1\%)$ as that for RBL,
and is as $57.2\%(53.7\%)$ as that for PCA.
\end{itemize}
The fist discovery should be explained that the Fourier spectra of the ramp function,
though their energy is mainly distributed in the low-frequency domain,
are indeed nonzero in the high-frequency domain and do not have a critical value $f_{c}$.
Therefore, a traffic matrix in the "ramp" dataset does not exactly satisfy the R-TMDM model we established in Section 2,
and the SPCP-TFC baseline method shows a small decline in accuracy.
The second discovery indicates that SPCP-TFC could still achieve more accurate baseline estimations than RBL and PCA
when the periodic traffic patterns follow the ramp functions instead of the sine functions.

\section{Conclusions and Future Work}
In this paper, we propose a refined traffic matrix decomposition model,
and introduce a novel traffic matrix baseline scheme named SPCP-TFC based on this model.
SPCP-TFC extends SPCP by using new time-frequency constraint functions for the baseline traffic and the noise traffic, respectively.
We design the APG algorithm for SPCP-TFC, whose convergence rate is $O(1/k^{2})$.
Our baseline scheme is demonstrated by simulations under appropriate settings.
Using three distinct metrics for evaluation,
we show that SPCP-TFC outperforms the existing baseline schemes RBL and PCA,
because it (i) archives a significant lower NRMSE;
(ii) extracts more precise temporal characteristics;
and (iii) leads to a more accurate estimation on smoothness.
The sensitivity of our baseline scheme with different values of parameter $\beta$,
as well as the baseline accuracy under different noise levels are discussed.
In addition, we simulate periodic traffic patterns which do not follow the sine functions,
and our experiment results indicate that though the SPCP-TFC baseline method has a small decline in accuracy on this "ramp" dataset,
it still achieves relatively lower NRMSE than RBL and PCA.

One significant problem in anomaly detection is to identify the normal but non-stationary traffic patterns,
and remove it from the total traffic.
Since the baseline component captured by this scheme is quite precise,
the residual component (i.e. the anomaly traffic and the noise traffic) should be a reasonable input for anomaly detection.
In our future work, we plan to design a new anomaly detection algorithm based on SPCP-TFC.

\section*{Acknowledgements}
We would like to acknowledge the suggestions from Prof. Jinping Sun at Beihang University on the earlier version of this work.
We would also like to acknowledge the constructive comments from the anonymous reviewers.

\section*{Appendix. Proof of Proposition 1}
Proof: For any two points $(A',E',N')$ and $(A'',E'',N'')$ in
$\mathbb{R}^{T\times P}\times\mathbb{R}^{T\times P}\times\mathbb{R}^{T\times P} $, we have
\begin{equation}\label{Lipschitz}
\begin{split}
        & \left\|\nabla f(A',E',N') - \nabla f(A'',E'',N'')\right\|_{F}^{2} \\
=      & \left\|[I +\beta W_{\mathrm{H}}(W_{\mathrm{H}})^{\top}](A'-A'') + (E'-E'') + (N'-N'')\right\|_{F}^{2}  + 2\left\|(A'-A'') + (E'-E'') + (N'-N'')\right\|_{F}^{2} \\
\leq & \Big{(} \beta\left\|W_{\mathrm{H}}(W_{\mathrm{H}})^{\top}(A'-A'')\right\|_{F}  + \|A'-A''\|_{F} + \|E'-E''\|_{F}  + \|N'-N''\|_{F} \Big{)}^{2} + 2 \Big{(}\|A'-A''\|_{F} + \|E'-E''\|_{F} + \|N'-N''\|_{F} \Big{)}^{2}\\
=      & \beta^{2}\left\|W_{\mathrm{H}}(W_{\mathrm{H}})^{\top}(A'-A'')\right\|_{F}^{2} +
        2\beta\left\|W_{\mathrm{H}}(W_{\mathrm{H}})^{\top}(A'-A'')\right\|_{F}
        \Big{(} \|A'-A''\|_{F}  + \|E'-E''\|_{F} + \|N'-N''\|_{F} \Big{)}  + \\
        & 6 \Big{(}\|A'-A''\|_{F}\|E'-E''\|_{F} +  \|A'-A''\|_{F}\|N'-N''\|_{F} + \|E'-E''\|_{F}\|N'-N''\|_{F} \Big{)} + \\
        & 3 \Big{(}\|A'-A''\|_{F}^{2} + \|E'-E''\|_{F}^{2} + \|N'-N''\|_{F}^{2} \Big{)} \\
\leq & (3\beta + \beta^{2})\left\|W_{\mathrm{H}}(W_{\mathrm{H}})^{\top}(A'-A'')\right\|_{F}^{2} +  (9+\beta)\Big{(}\|A'-A''\|_{F}^{2} + \|E'-E''\|_{F}^{2} + \|N'-N''\|_{F}^{2} \Big{)}.\\
\end{split}
\end{equation}
As $W_{\mathrm{H}}(W_{\mathrm{H}})^{\top}$ is a projection operator in $\mathbb{R}^{T}$,
for each $p\in\{1,...,P\}$,
\begin{equation}
\left\|W_{\mathrm{H}}(W_{\mathrm{H}})^{\top}(A'_{p}-A''_{p}) \right\| \leq \left\|A'_{p}-A''_{p}\right\|.
\end{equation}
Therefore, we have
\begin{equation}\label{projection_property}
\begin{split}
           \left\|W_{\mathrm{H}}(W_{\mathrm{H}})^{\top}(A'-A'')\right\|_{F}^{2}
=        & \sum_{p=1}^{P}\left\|W_{\mathrm{H}}(W_{\mathrm{H}})^{\top}(A'_{p}-A''_{p}) \right\|^{2} \\
\leq   & \sum_{p=1}^{P}\left\|A'_{p}-A''_{p}\right\|^{2} \\
=        &\|A'-A''\|_{F}^{2}.
\end{split}
\end{equation}
Add (\ref{projection_property}) to ($\ref{Lipschitz}$) and let $L = \sqrt{9 + 4\beta + \beta^{2}}$, we have
\begin{equation}
\begin{split}
        \left\|\nabla f(A',E',N') - \nabla f(A'',E'',N'')\right\|_{F}^{2}
\leq & (9 + 4\beta + \beta^{2}) \left(\left\|A'-A''\right\|_{F}^{2} + \left\|E'-E''\right\|_{F}^{2} + \left\|N'-N''\right\|_{F}^{2} \right) \\
=     & L \left\|(A',E',N') - (A'',E'',N'')\right\|_{F}^{2}. \\
\end{split}
\end{equation}
\noindent This completes the proof. $\square$


\begin{thebibliography}{00}

\bibitem{Lakhina_2004}
A. Lakhina, K. Papagiannaki, M. Crovella, C. Diot, E. D. Kolaczyk, and N. Taft.
Structural analysis of network traffic flows.
Proc. ACM SIGMETRICS, 2004.

\bibitem{Lakhina_sigcomm}
A. Lakhina, M. Crovella, and C. Diot.
Diagnosing Network-Wide Traffic Anomalies.
Proc. ACM SIGCOMM, 2004.

\bibitem{Bandara_2011}
V. W. Bandara and A. P. Jayasumana.
Extracting Baseline Patterns in Internet Traffic Using Robust Principal Components.
Proc. IEEE LCN, 2011.

\bibitem{Wang_2012}
Z. Wang, K. Hu, K. Xu, B. Yin and X. Dong.
Structural Analysis of Network Traffic Matrix via Relaxed Principal Component Pursuit.
Computer Networks, vol. 56, no. 7, pp. 2049-2067, 2012.

\bibitem{Hajji_2003}
H. Hajji.
Baselining Network Traffic and Online Faults Detection.
Proc. IEEE ICC, 2003.

\bibitem{Conti_2010}
P. Conti, L. Giovanni and M. Naldi.
Blind Maximum likelihood estimation of Traffic Matrices under Long-Range Dependent Traffic.
Computer Networks, vol. 54, no. 15, pp. 2626-2639, 2010.

\bibitem{Nie_2013}
L. Nie, D. Jiang, and L. Guo.
A power laws-based reconstruction approach to end-to-end network traffic.
Journal of Network and Computer Applications, vol. 36 no. 3, pp. 898-907, 2013.

\bibitem{Willinger_2011}
Walter Willinger.
On a new Internet Traffic Matrix (Completion) Problem.
Duke Workshop on Sensing and Analysis of High-Dimensional Data (SAHD), 2011.
http://sahd.pratt.duke.edu/2011\_files/Videos\_and\_Slides.html.

\bibitem{Cisco_2007}
Cisco Systems.
NetFlow Services Solutions Guide, 2007.
http://www.cisco.com/en/US/docs/ios/solutions\_docs/netflow/nfwhite.pdf.

\bibitem{Rubinstein_2009}
B. Rubinstein, B. Nelson, L. Huang, A. Joseph, S. Lau, S. Rao,
N. Taft and J. Tygar. ANTIDOTE: understanding and defending
against poisoning of anomaly detectors. Proc. ACM IMC, 2009.

\bibitem{Ma_RPCA1}
E. Candes, X. Li, Y. Ma, and J. Wright. Robust principal component analysis?
Journal of the ACM. vol. 58, no. 3, pp. 1-37, 2011.

\bibitem{Abdelke_2010}
A. Abdelke, Y. Jiang, W. Wang, A. Aslebo, and O. Kvittem.
Robust traffic anomaly detection with principal component pursuit.
Proc. ACM CoNEXT Student Workshop, 2010.

\bibitem{Soule_2007}
A. Soule, A. Nucci, R. L. Cruz, E. Leonardi, N. Taft.
Estimating dynamic traffic matrices by using viable routing changes.
IEEE/ACM Transactions on Networking. vol. 15, no.3, pp. 485-498, 2007.

\bibitem{Zhou_2010}
Z. Zhou, X. Li, J. Wright, E. Candes, and Y. Ma.
Stable principal component pursuit.
Proc. IEEE ISIT, 2010.

\bibitem{Barford_2002}
P. Barford, J. Kline, D. Plonka and A. Ron.
A Signal Analysis of Network Traffic Anomalies.
Proc. ACM IMW, 2002.

\bibitem{Roughan_2002}
M. Roughan and J. Gottlieb.
Large-scale Measurement and Modeling of Backbone Internet Traffic.
Proc. SPIE ITCom, 2002.

\bibitem{Kokoszka_2000}
P. Kokoszka and T. Mikosch.
The periodogram at the Fourier frequencies.
Stochastic Processes and their Applications, vol. 86, no. 1, pp. 49-79, 2000.

\bibitem{Lin_2009}
Z. Lin, A. Ganesh, J. Wright, L. Wu, M. Chen and Y. Ma.
Fast convex optimization algorithms for exact recovery of a corrupted low-rank matrix.
Proc. IEEE CAMSAP, 2009.

\bibitem{Combettes_2010}
P. L. Combettes and J.-C. Pesquet.
Proximal splitting methods in signal processing.
Fixed-Point Algorithms for Inverse Problems in Science and Engineering,
pp. 185-212. Springer, New York, 2011.

\bibitem{Beck_2009}
A. Beck and M. Teboulle.
A fast iterative shrinkage-thresholding algorithm for linear inverse problems.
SIAM Journal on Imaging Sciences, vol. 2, no. 1, pp. 183-202, 2009.

\bibitem{Roughan_2005}
M. Roughan.
Simplifying the synthesis of Internet traffic matrices.
Computer Communication Review, vol. 35, no. 5, pp. 93-96, 2005.

\end{thebibliography}
\end{document}